\documentclass[trackchanges,twocolumn]{aastex7}
\received{June 23, 2025}
\accepted{July 14, 2025}
\submitjournal{ApJL}
\begin{document}

\title{A Transiting Giant on a 7.7-Year Orbit Revealed by TTVs in the TOI-201 System}

\author[orcid=0000-0002-4195-5781,gname=Gracjan,sname=Maciejewski]{Gracjan Maciejewski}
\affiliation{Institute of Astronomy, Faculty of Physics, Astronomy and Informatics, Nicolaus Copernicus University in Toru\'n, Grudziadzka 5, 87-100 Toru\'n, Poland}
\email[show]{gmac@umk.pl}  

\author{Weronika \L{}oboda} 
\affiliation{Institute of Astronomy, Faculty of Physics, Astronomy and Informatics, Nicolaus Copernicus University in Toru\'n, Grudziadzka 5, 87-100 Toru\'n, Poland}
\email{wika.loboda@gmail.com}

\begin{abstract}
We report the detection and characterization of TOI-201~c, a long-period transiting companion to the warm Jupiter TOI-201~b. Its presence was first inferred from high-amplitude transit timing variations (TTVs) in TOI-201~b, pointing to a massive outer body on a $7.7^{+1.0}_{-0.6}$-year eccentric orbit. This prediction was confirmed when TESS observed a transit of TOI-201~c, precisely constraining its orbital geometry. A joint fit to TTVs, transit photometry, and archival radial velocities yields a mass of $14.2^{+1.0}_{-1.2}$~$M_{\rm Jup}$ and an eccentricity of $0.643^{+0.009}_{-0.021}$. The mutual inclination between planets~b and~c is $2.9^{+4.8}_{-4.4}$ degrees, indicating a nearly coplanar architecture. Long-term numerical integrations confirm dynamical stability over gigayear timescales and predict that transits of TOI-201~b will cease within a few thousand years. TOI-201~c ranks among the longest-period transiting planets with well-constrained properties. Its detection via TTVs, followed by a confirmed transit, represents a rare observational sequence and highlights the power of TTVs and photometric monitoring to uncover distant companions. The TOI-201 system offers a valuable laboratory for testing models of giant planet formation, migration, and secular evolution in multi-planet systems.
\end{abstract}



\section{Introduction}\label{sec:Intro}

The star TOI-201 (HD~39474, HIP~27515) is an F6--7 dwarf \citep{1975mcts.book.....H,2021AJ....161..235H}. Two candidate transiting planets were identified in its light curve, obtained by the \textit{Transiting Exoplanet Survey Satellite} \citep[TESS;][]{2015JATIS...1a4003R}, and reported by \citet{2021ApJS..254...39G}. The deeper signal corresponds to TOI-201~b, showing a transit-like feature with a period $P_{\rm b} \approx 53$~d, depth $\approx$6~ppth, and duration $\approx$280~minutes \citep{2020MNRAS.498.1726M,2021ApJS..254...39G,2021ApJS..255....6D}. \citet{2021AJ....161..235H} confirmed its planetary nature and determined a mass of $M_{\rm b} = 0.42^{+0.05}_{-0.03} \, M_{\rm Jup}$ and a radius of $R_{\rm b} = 1.008^{+0.012}_{-0.015} \, R_{\rm Jup}$, with an eccentric orbit ($e_{\rm b} = 0.28^{+0.06}_{-0.09}$) and semi-major axis $a_{\rm b} = 0.30^{+0.02}_{-0.03}$~au. In addition to the planet's induced radial velocity signal, \citet{2021AJ....161..235H} also reported a long-term trend, initially attributed to stellar activity based on its complexity and the star's estimated young age of $0.87^{+0.46}_{-0.49}$~Gyr.

The second candidate, TOI-201.02, may be a $1.7 \, R_{\oplus}$ planet in a 5.8-day orbit \citep{2021ApJS..254...39G}, though its planetary status remains unconfirmed by follow-up observations \citep{2021AJ....161..235H}.

Thanks to its long orbital period and coverage across multiple TESS sectors, TOI-201~b was included in our program to search for short-period transit timing variations (TTVs), which may signal the presence of exomoons perturbing their host planets \citep{2021MNRAS.500.1851K}. Transits observed during TESS Cycles~1 and~3 were consistent with a linear ephemeris. However, this changed with the inclusion of three transits from Cycle~5, which revealed an apparent increase in TOI-201~b's orbital period \citep{2024Loboda}. This unexpected deviation prompted a deeper investigation of the system, ultimately leading to the detection of an additional massive, transiting companion in a wider orbit---a discovery we present in this Letter.

\section{TESS Observations} \label{sec:Obs}

TOI-201 lies approximately $12^{\circ}$ from the southern ecliptic pole, near the edge of the TESS Southern Continuous Viewing Zone. As a result, the system was monitored almost continuously during Cycles~1, 3, and the first half of Cycle~5. To date, it has been observed in 32 sectors with a two-minute cadence, yielding complete transits of TOI-201~b in 15 of them.

We used the \texttt{Lightkurve v2.0} package \citep{2018ascl.soft12013L} to extract light curves via aperture photometry from the two-minute cadence target pixel files. For sectors contaminated by background scattered light, we applied the \texttt{RegressionCorrector} module to enhance photometric precision. Transits were masked using 15-minute buffers around visually refined mid-transit times. To remove instrumental systematics and stellar variability on timescales from days to a month, we applied a Savitzky--Golay filter \citep{1964AnaCh..36.1627S} with a 12-hour window, resulting in flattened, normalized light curves suitable for transit analysis.

\section{Transit modeling} \label{sec:TrModels}

We employed a customized version of the \texttt{Transit Analysis Package} \citep[\texttt{TAP};][]{2012AdAst2012E..30G} to derive key transit parameters: the mid-transit time $T_{\rm mid}$, planet-to-star radius ratio $R/R_{\star}$, transit chord duration $\tau$ (defined as the interval between the moments when the center of the planet crosses the stellar limb at ingress and egress), and the impact parameter $b$ (normalized to $R_{\star}$). The stellar limb darkening was modeled using a quadratic law, with coefficients $u_{\rm 1,TESS}$ and $u_{\rm 2,TESS}$ interpolated from the tables of \citet{2011AA...529A..75C} using the stellar parameters from \citet{2021AJ....161..235H}. Gaussian priors with widths of 0.1 and 0.2 were applied, following \citet{2022AJ....163..228P}. Residual baseline trends near each transit were removed using second-order polynomials.

\texttt{TAP} was applied to light curve segments centered on each transit, extracted with margins of $2.5\,\tau$. In the first fitting pass, all parameters except $T_{\rm mid}$ were linked across transits to determine global values. In a second iteration, we allowed $R/R_{\star}$, $\tau$, and $b$ to vary per transit to check for potential variability. No significant inter-transit differences were found. In an additional test, $u_{\rm 1,TESS}$ and $u_{\rm 2,TESS}$ were treated as free parameters. While the fitted values remained consistent with theoretical predictions within $1\sigma$, their uncertainties increased by a factor of two. We therefore adopted the literature values for our final modeling.

\begin{figure}[ht!]
\includegraphics[width=1.0\columnwidth]{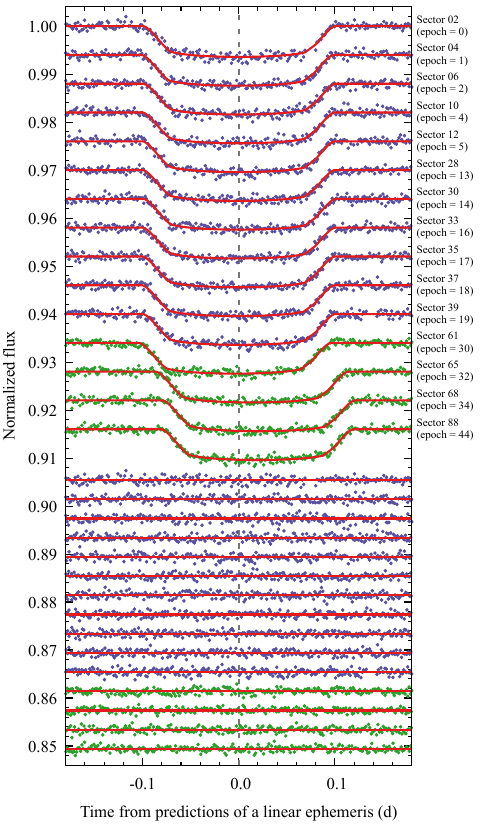}
\caption{Transit light curves of TOI-201~b observed with TESS. Transits consistent with a linear ephemeris are shown in blue, with predicted mid-transit times marked by the dashed gray line. Transits exhibiting significant timing offsets, particularly those from Sectors~65--88, are plotted in green. Best-fitting models are overlaid in red, with residuals shown below.}
\label{fig:201bTransits}
\end{figure}

We explored the parameter space using ten MCMC walkers, each run for $10^6$ steps with a 10\% burn-in. The chain convergence was verified with the multivariate
Gelman-Rubin statistic and the number of effective samples. Median values and uncertainties were taken from the posterior distributions at the 50th, 15.9th, and 84.1st percentiles. The average transit parameters are listed in Table~\ref{tab:sys_pars}, while individual transit fits are summarized in Table~\ref{tab:ttimes}. The transit light curves of TOI-201~b, along with best-fitting models and residuals, are shown in Fig.~\ref{fig:201bTransits}.

\begin{figure}[ht!]
\includegraphics[width=1.0\linewidth]{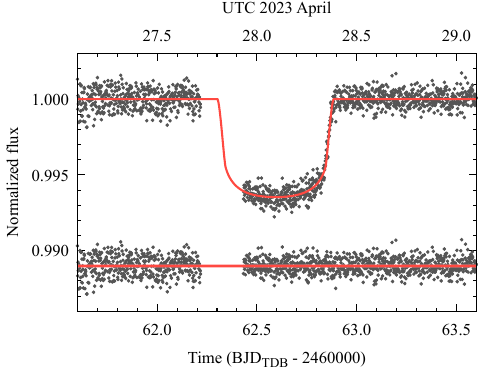}
\caption{Transit light curve of TOI-201~c detected in TESS Sector~64. The best-fitting model is shown in red, with residuals plotted below.}
\label{fig:201cTransits}
\end{figure}

Our visual inspection of Sector~64 revealed an incomplete transit inconsistent with TOI-201~b. Although similar in depth, it exhibited a significantly longer duration---estimated at 10--15 hours. The ingress was missed due to a data gap during the apogee downlink in TESS orbit~136. We attribute this event to a newly identified outer companion, TOI-201~c. This transit was modeled using the same \texttt{TAP} procedure. The resulting fit and residuals are presented in Fig.~\ref{fig:201cTransits}, with parameters included in Table~\ref{tab:sys_pars}.

\section{Two-planetary Dynamical Model} \label{sec:Dynamical Model}

To investigate potential gravitational interactions in the TOI-201 system, we analyzed the transit timing behavior of TOI-201~b. Our analysis began by refining the planet's linear ephemeris using transits from TESS Sectors~2--39. We adopted the standard form:
\begin{equation}
 \label{Eq:ephem}
  T_{\rm mid,b}(E) = T_{\rm 0,b} + P'_{\rm b} \times E \, ,
\end{equation}
where $E$ is the transit number counted from a reference epoch. We determined $T_{\rm 0,b} = 2458376.05207 \pm 0.00031$~$\rm BJD_{TDB}$ (corresponding to the first TESS-observed transit) and an apparent orbital period of $P'_{\rm b} = 52.978199 \pm 0.000025$~d. These values were obtained by sampling the posterior probability distribution using an MCMC analysis with 100 chains, each consisting of $10^4$ steps and a 10\% burn-in.

\begin{figure*}[ht!]
\includegraphics[width=1.0\linewidth]{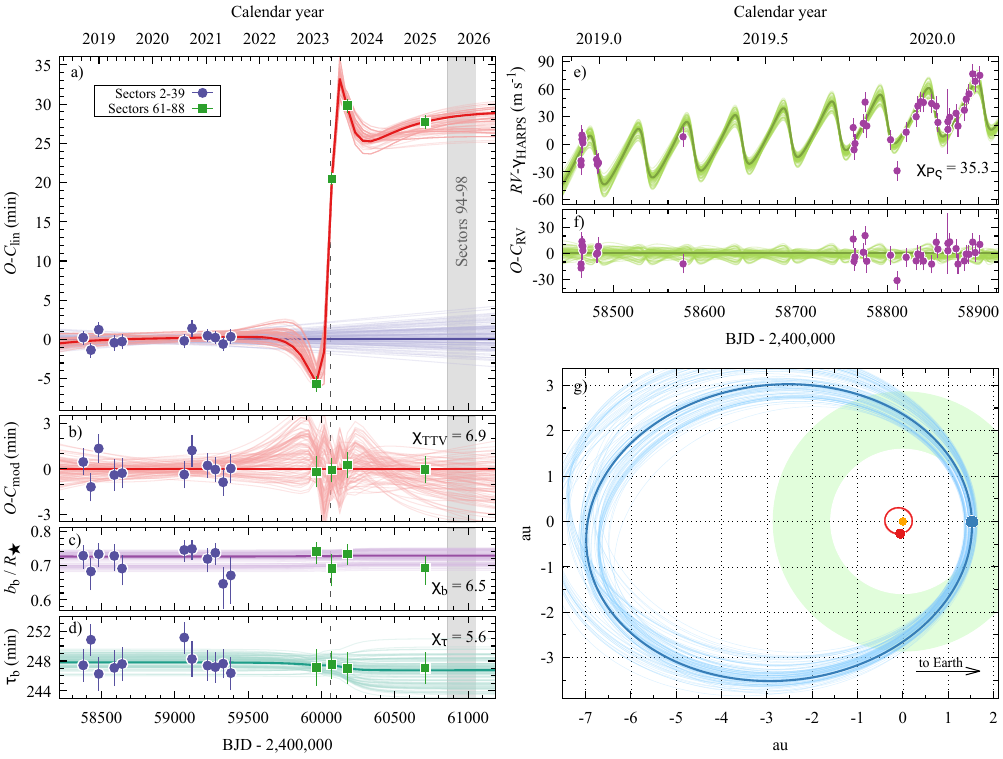}
\caption{Transit observables for TOI-201~b from the dynamical model. Panel~a: Transit timing residuals relative to the refined linear ephemeris, derived using mid-transit times from TESS Sectors~2--39 (blue dots). The thick blue line represents the linear ephemeris propagated to later epochs, where deviations caused by TOI-201~c become apparent. A set of 50 posterior realizations is overplotted to illustrate the uncertainty in this ephemeris. Mid-transit times of TOI-201~b that deviate significantly from the linear prediction are highlighted in green. The best-fitting two-planet dynamical model is shown as a thick red line, with 100 posterior samples overplotted to illustrate model uncertainty. The shaded gray region indicates the window of upcoming TESS observations. The vertical dashed line marks the transit of TOI-201~c. Panel~b: Transit timing residuals relative to the best-fitting dynamical model. Panels~c and~d: Evolution of the impact parameter $b_{\rm b}$ and transit duration $\tau_{\rm b}$, respectively, as predicted by the model. Uncertainties are indicated by the ensemble of posterior realizations. Panel~e: HARPS RV measurements with the best-fitting two-planet dynamical model overplotted. The RV modulation induced by TOI-201~b is superimposed on a long-term trend caused by TOI-201~c. The residuals are shown in panel~f. Panel g: schematic diagram showing the astrocentric architecture of the TOI-201 system. The orbits of planets~b and~c are shown in red and blue, respectively. Uncertainties are illustrated using 100 posterior samples. Planet positions at the epoch of $T_{\rm mid,c}$ are marked with dots, and the host star is shown as an orange dot. Sizes of all bodies are not to scale. The pale green region indicates the circumstellar habitable zone, using conservative inner and outer boundaries (runaway and maximum greenhouse) computed via Equations~(2) and~(3) from \citet{2013ApJ...765..131K}.}
\label{fig:201bTiming}
\end{figure*}

Figure~\ref{fig:201bTiming}, panel~a, shows the timing residuals relative to this refined linear ephemeris. Transits from Sectors~2--39, which align well with the linear model, are marked in blue, while later transits---those from Sectors~61 and 65--88---exhibit significant deviations of up to 30 minutes and are shown in green. A bundle of 50 posterior realizations of the linear ephemeris is overplotted in blue to illustrate its propagated uncertainty. These deviations from linearity are too large to be attributed to measurement noise and strongly suggest the presence of dynamical perturbations from a second body in the system.

Motivated by these anomalies, we constructed a two-planet dynamical model using the \texttt{TTVFast} integrator \citep{2014ApJ...787..132D} and the dynamic nested sampling algorithm from the \texttt{dynesty} package \citep{2020MNRAS.493.3132S}. The model simultaneously fits TTVs, transit shape parameters, and radial velocity (RV) data. Our observational inputs included the mid-transit times, durations ($\tau_{\rm b}$), and impact parameters ($b_{\rm b}$) for 15 observed transits of TOI-201~b, as well as single measurements of $T_{\rm mid,c}$, $\tau_{\rm c}$, and $b_{\rm c}$ for TOI-201~c. These were supplemented with 39 HARPS RV measurements from \citet{2021AJ....161..235H}, chosen for their superior precision and longer time coverage (from November~2018 to August~2020). Other RV datasets were not included in the final fit due to their lower quality, with uncertainties comparable to or exceeding the RV amplitude induced by TOI-201~b. We note that including these additional data leaves our results essentially unchanged.

Assuming Gaussian and uncorrelated errors on all input observables, the likelihood function took the form:
\begin{equation}
 \label{Eq:logL}
  \log \mathcal{L} = -\frac{1}{2} \sum_i \left[ \frac{(\mathcal{O}_i - \mathcal{M}_i)^2}{\sigma_i^2} + \log(2\pi\sigma_i^2) \right]  \, , \;
\end{equation}
where $\mathcal{O}_i$ and $\sigma_i$ are the observed values and uncertainties, and $\mathcal{M}_i$ are the corresponding model predictions.

The dynamical model included the following free parameters: the masses $M_{\rm b}$ and $M_{\rm c}$, orbital periods $P_{\rm b}$ and $P_{\rm c}$, eccentricities $e_{\rm b}$ and $e_{\rm c}$, inclinations $i_{\rm b}$ and $i_{\rm c}$, arguments of periastron $\omega_{\rm b}$ and $\omega_{\rm c}$, and mean anomalies $\theta_{\rm m, b}$ and $\theta_{\rm m, c}$ at a common epoch. The longitude of the ascending node for TOI-201~b was fixed to $0^{\circ}$, establishing the reference for TOI-201~c's relative node, $\Omega_{\rm c}$. The model also included RV offset and jitter terms. The stellar mass was fixed at $M_{\star} = 1.316 \, M_{\odot}$ from \citet{2021AJ....161..235H}, while the stellar radius $R_{\star}$ was treated as a free parameter to allow consistent derivation of $b$ and $\tau$ from dynamical quantities. Physical consistency was enforced by requiring agreement between the stellar density inferred from each planet's transit and the value derived from the fixed $M_{\star}$ and fitted $R_{\star}$.

In addition to predicting transit times and radial velocities, \texttt{TTVFast} also computes the sky-projected astrocentric distance $r_{\rm sky}$ and the sky-projected astrocentric velocity $v_{\rm sky}$, which can be converted into $b$ and $\tau$ via the following relations:
\begin{equation}
  \label{Eq:get_b}
  b = \frac{r_{\rm sky}}{R_{\star}} \, ,
\end{equation}
\begin{equation}
  \label{Eq:get_tau}
  \tau = \frac{2}{v_{\rm sky}} \sqrt{R_{\star}^2 - r_{\rm sky}^2} \, .
\end{equation}
These expressions link the modeled orbital geometry to observable transit parameters, adding further constraints on the orbital evolution.

We explored the parameter space with multiple runs of dynamic nested sampling, each using 6400 live points. Convergence was reached when the change in log-evidence dropped below $\Delta \ln \mathcal{Z} < 0.001$. To avoid trapping in local maxima, prior grids were imposed on $P_{\rm c}$ (spanning 2350--9850~d in steps of 200--500~d) and $\Omega_{\rm c}$ (spanning $0^{\circ}$--$330^{\circ}$ in $30^{\circ}$ intervals, each within $\pm 20^{\circ}$). The inclination $i_{\rm c}$ was explored in two separate regimes: $i_{\rm c} \le 90^\circ$ and $i_{\rm c} \ge 90^\circ$, while $i_{\rm b}$ was restricted to $i_{\rm b} \le 90^\circ$ to avoid degeneracy in $i_{\rm b}$ and $\tau_{\rm b}$.

The resulting log-likelihood surface showed clear, distinct maxima for $P_{\rm c}$ in the 2600--3300~d range, with solutions most strongly clustered around $\Omega_{\rm c} \approx 0^{\circ}$, $90^{\circ}$, $180^{\circ}$, and $270^{\circ}$. Among these, configurations near $\Omega_{\rm c} \approx 0^{\circ}$ were decisively favored by the Bayesian Information Criterion (BIC), with $\Delta \mathrm{BIC} > 15$ relative to other maxima. The inclination of planet c exhibited a bimodal distribution with distinct peaks at $i_{\rm c}$ and $180^\circ - i_{\rm c}$. Although both families of solutions fit the data equally well (with $\Delta \mathrm{BIC} < 2$), the best-fitting model lies within the $i_{\rm c} < 90^\circ$ regime, and we focus on that branch in the following discussion. 

The best-fitting model parameters are reported in Table~\ref{tab:sys_pars}, with uncertainties defined by the 15.9th and 84.1st percentiles of merged posterior samples from all models within $\Delta \mathrm{BIC} < 2$ of the best solution. The convergence of this merged sample was verified qualitatively with visual inspection of the profile likelihood for each model parameter \citep[e.g.,][]{2021PSJ.....2....1A}, which confirmed that the posterior modes correspond to isolated, well-resolved regions of parameter space. TOI-201~c emerges as a super-Jupiter or low-mass brown dwarf ($M_{\rm c} = 14.2^{+1.0}_{-1.2} \, M_{\rm Jup}$) in a highly eccentric ($e_{\rm c} = 0.643^{+0.009}_{-0.021}$), long-period ($P_{\rm c} = 7.7^{+1.0}_{-0.6}$ years) orbit. The mutual inclination between the planetary orbits is $i_{\rm bc} = 2.9^{+4.8}_{-4.4}$ degrees, consistent with a coplanar configuration.

Figure~\ref{fig:201bTiming}, panel~a, shows the dynamical model's fit to the transit timing residuals (red line) and 100 posterior realizations. The corresponding residuals are plotted in panel~b. Panels~c and d show the predicted evolution of $b_{\rm b}$ and $\tau_{\rm b}$, respectively. The model predicts only a 0.4\% decrease in $b_{\rm b}$ during the 2023 periastron passage---below current detection limits---and a small, marginally significant change in $\tau_{\rm b}$ of approximately 1 minute. The observed difference in $\tau_{\rm b}$ between pre- and post-2023 transits is $0.5 \pm 2.4$ minutes, consistent with this prediction. Panels~e and~f show the RV component of the model and its residuals, respectively, while panel~g presents a schematic diagram of the system architecture.

Future TESS observations (Sectors~94--98) are expected to include four additional transits of TOI-201~b. These will reduce uncertainties on $\tau_{\rm b}$ by an estimated 30\%, but likely still fall short of definitively resolving the predicted signal.

\section{Discussion} \label{sec:Discussion} 

TOI-201~c is among the longest-period transiting exoplanets discovered to date, with an orbital period comparable to that of K2-311~b \citep{2018A&A...615L..13G}, which orbits an evolved subgiant. Notably, TOI-201~c is the longest-period transiting planet discovered by TESS. What distinguishes it is not only its extreme orbital period, but also the tight constraints on its orbital and physical parameters. Its coexistence with TOI-201~b in a nearly coplanar, dynamically interacting configuration makes the TOI-201 system a rare and valuable laboratory for testing theories of planetary formation, migration, and long-term dynamical evolution.

With a mass of $\sim$$14\,M_{\rm Jup}$, TOI-201~c lies near the deuterium-burning threshold traditionally used to distinguish giant planets from brown dwarfs. Its classification depends critically on formation history: if it formed via core accretion, as suggested by its $\sim$4~au orbit and the host star's high metallicity (${\rm [Fe/H]} = 0.240 \pm 0.036$; \citealt{2021AJ....161..235H}), it would be considered a planet \citep{2018ApJ...853...37S}. Its bulk density of $\sim$19~${\rm g\,cm^{-3}}$ places it firmly on the empirical mass--density sequence of massive exoplanets \citep{2015ApJ...810L..25H}, reinforcing its classification as a high-mass giant planet rather than a stellar brown dwarf.

Despite its long orbital period, the extreme eccentricity of TOI-201~c brings it to within 1.5~au of its host star at periastron, leading to strong irradiation-driven variations over the orbit. Assuming a Bond albedo of 0.3 and full energy redistribution \citep{1999ApJ...513..879M}, the planet's orbit-averaged equilibrium temperature \citep{2011ApJ...726...82C} is $169 \pm 6$~K. However, incident stellar flux swings from 0.05 to 1.1~$S_{\odot}$, periodically bringing the planet into the habitable zone at periastron (see panel g of Fig.~\ref{fig:201bTiming}).

We explored the long-term dynamical evolution of the best-fitting model using the \texttt{SWIFT} symplectic integrator \citep{1994Icar..108...18L, 2013ascl.soft03001L}, running numerical simulations over 50 kyr with a timestep of 1\% of $P_{\rm b}$. To assess orbital stability, we computed the chaos indicator $D$ \citep{1990Icar...88..266L,1993PhyD...67..257L} for each planet, based on frequency diffusion across two 25-kyr intervals \citep[e.g.,][]{2005A&A...440..751C}. The resulting values, $\log D_{\rm b} = -7.0$ and $\log D_{\rm c} = -10.3$, indicate long-term stability on Gyr timescales. Indeed, in a separate 800-Myr integration---comparable to the system's estimated age---both planetary orbits remained regular, with no signs of chaotic evolution.

The orbital elements of TOI-201~b show secular modulation under the perturbation of TOI-201~c. Its eccentricity oscillates between 0.12 and 0.32 on a $\sim$25 kyr timescale, while the inclination cycles between $87^\circ$ and $93^\circ$ over $\sim$21 kyr, exhibiting a step-like evolution pattern, triggered by gravitational perturbations from TOI-201~c during its periastron passages. These short-timescale variations in $i_{\rm b}$ induce corresponding discrete changes in $\tau_{\rm b}$. While these individual steps are small---about 1 minute---they accumulate across successive orbits, leading to measurable long-term trends. Transit visibility is likewise modulated: TOI-201~b remains observable during only $\sim$30\% of its inclination cycle, with transits expected to cease around the year 3000 and resume approximately 7000 years later. In contrast, transits of TOI-201~c persist throughout its secular cycle, with $b_{\rm c}$ varying between 0.35 and 0.8 and durations oscillating between 10 and 15 hours. Although TOI-201~c is predicted to exhibit smooth TTVs with an amplitude of approximately 4 days, these occur on a timescale of $\sim$3000 years and will remain undetectable in the foreseeable future.

Our dynamical model revises the mass of TOI-201~b upward by approximately 40\% compared to the value reported by \citet{2021AJ....161..235H}, a discrepancy significant at the $\sim$$2\sigma$ level. The difference likely stems from Gaussian process detrending in their adopted model~6, which may partially absorb the RV signal. Our derived RV semi-amplitude $K_{\rm b}$ agrees more closely with values from their alternative fits: model~3 (single planet, eccentric orbit) and model~5 (eccentric orbit with a quadratic trend).

We also explored the dynamical role of TOI-201.02, a candidate inner planet detected via transits. Using the mass--radius relation recently refined by \citet{2024A&A...686A.296M}, we estimate its mass to be $\sim$$5\,M_{\oplus}$. We introduced a planet of this mass into our dynamical model to assess its interactions with the two confirmed giants. The test body remains dynamically decoupled from TOI-201~c on decadal timescales, but experiences TTVs induced by TOI-201~b, with a characteristic period of $\sim$900 days and amplitudes reaching $\sim$5 minutes. These TTVs would be challenging to detect due to the planet's shallow transit depth and current observational limitations. Conversely, the gravitational influence of TOI-201.02 on TOI-201~b results in TTVs of less than $\sim$20 seconds, well below existing detection thresholds.

\section{Conclusions} \label{sec:Conclusions}

We presented a comprehensive dynamical model of the TOI-201 system, based on a joint analysis of transit timing variations, radial velocity measurements, and the newly observed transit of the outer companion. The detection of TOI-201~c was initially inferred from the high-amplitude TTV signal it induces on TOI-201~b---an uncommon case in which a long-period transiting planet was first revealed through its dynamical imprint. Its subsequent transit detection by TESS confirms the prediction and provides a crucial observational anchor for the system's architecture.

TOI-201~c follows an eccentric $\sim$8-year orbit and has a mass of $\approx14\,M_{\rm Jup}$, placing it near the boundary between massive planets and brown dwarfs. Its nearly coplanar orbit with TOI-201~b, along with the high metallicity of the host star, supports a formation pathway via core accretion. Long-term numerical integrations confirm the system's stability over gigayear timescales and predict that transits of TOI-201~b will vanish within the next few millennia due to secular oscillations in orbital inclination.

As one of the longest-period transiting planets with precisely constrained properties, TOI-201~c provides a rare opportunity to probe the architecture, formation, and long-term evolution of giant planetary systems. TOI-201 thus emerges as a valuable natural laboratory for testing theories of planet--planet interactions, orbital dynamics, and the transition between planetary and substellar regimes. Continued observations---particularly with future TESS coverage and radial velocity monitoring---will further refine our understanding and may reveal additional dynamical features in this remarkable system.

\begin{deluxetable*}{llcccl}
\tablewidth{0pt}
\tablecaption{Systemic parameters for TOI-201 \label{tab:sys_pars}}
\tablehead{
& \colhead{Parameter} & \colhead{Units} & \colhead{Prior} & \colhead{Value} & \colhead{Source}
}
\startdata
\multicolumn{6}{l}{TOI~201~b}\\
 & Orbital period, $P_{\rm{b}}$ & d & $\mathcal{U}(52.96,52.99)$ & $52.977946^{+0.000031}_{-0.000040}$ & TTVFast dynamical model\\
 & Orbital eccentricity, $e_{\rm{b}}$ & $-$ & $\mathcal{U}(0.0,0.5)$ & $0.318^{+0.020}_{-0.022}$ & TTVFast dynamical model\\
 & Orbital inclination, $i_{\rm{b}}$ & $^{\circ}$ & $\mathcal{U}(88,90)$ & $88.747^{+0.029}_{-0.032}$ & TTVFast dynamical model\\
 & Argument of periastron, $\omega_{\rm{b}}$ & $^{\circ}$ & $\mathcal{U}(0,360)$ & $81.3^{+4.4}_{-5.6}$ & TTVFast dynamical model\\
 & Longitude of the ascending node, $\Omega_{\rm{b}}$ & $^{\circ}$ & fixed & $0.0$ & TTVFast dynamical model\\
 & Mean anomaly, $\theta_{\rm m, b}$ & $^{\circ}$ & $\mathcal{U}(180,360)$ & $323.2^{+2.8}_{-2.3}$ & TTVFast dynamical model\\
 & Radius ratio, $R_{\rm{b}}/R_{\star}$ & $-$ & $-$ & $0.07919\pm{0.00028}$ & TAP transit model \\
 & Transit impact parameter, $b_{\rm{b}}$ & $-$ & $-$ & $0.729^{+0.011}_{-0.010}$ & TAP transit model\\
 & Transit duration, $\tau_{\rm{b}}$ & min & $-$ & $247.16^{+0.89}_{-0.87}$ & TAP transit model\\
 & Total transit duration, $\tau_{\rm{14, b}}$ & min & $-$ & $287.4\pm1.2$ & derived\\
 & Mass, $M_{\rm{b}}$ & $M_{\rm{Jup}}$ & $\mathcal{U}(0.1,0.9)$ & $0.580^{+0.070}_{-0.061}$ & TTVFast dynamical model\\
 & Radius, $R_{\rm{b}}$ & $R_{\rm{Jup}}$ & $-$ & $1.055^{+0.014}_{-0.009}$ & derived\\
 & Density, $\rho_{\rm{b}}$ & $\rm{g~cm}^{-3}$ & $-$ & $0.654^{+0.081}_{-0.074}$ & derived\\
 & Semi-major axis, $a_{\rm{b}}$ & au & $-$ & $0.3026\pm0.0021$ & derived\\
 & RV amplitude, $K_{\rm{b}}$ & $\rm{m~s}^{-1}$ & $-$ & $27.6^{+3.3}_{-2.9}$ & derived\\
\multicolumn{6}{l}{TOI~201~c} \\
& Orbital period, $P_{\rm{c}}$ & d & $\mathcal{U}(2350,9850)$ & $2800^{+360}_{-210}$ & TTVFast dynamical model\\
 & Orbital eccentricity, $e_{\rm{c}}$ & $-$ & $\mathcal{U}(0.5,0.9)$ & $0.643^{+0.009}_{-0.021}$ & TTVFast dynamical model\\
 & Orbital inclination, $i_{\rm{c}}$ & $^{\circ}$ & $\mathcal{U}(89,91)$ & $89.896^{+0.030}_{-0.017}$ & TTVFast dynamical model\\
 & Argument of periastron, $\omega_{\rm{c}}$ & $^{\circ}$ & $\mathcal{U}(0,360)$ & $95.2^{+3.2}_{-7.2}$ & TTVFast dynamical model\\
 & Longitude of the ascending node, $\Omega_{\rm{c}}$ & $^{\circ}$ & $\mathcal{U}(-20,350)$ & $2.7^{+5.2}_{-4.8}$ & TTVFast dynamical model\\
 & Mean anomaly, $\theta_{\rm m, c}$ & $^{\circ}$ & $\mathcal{U}(0,360)$ & $115^{+27}_{-19}$ & TTVFast dynamical model\\
 & Radius ratio, $R_{\rm{c}}/R_{\star}$ & $-$ & $-$ & $0.07449^{+0.00087}_{-0.00091}$ & TAP transit model\\
 & Transit impact parameter, $b_{\rm{c}}$ & $-$ & $-$ & $0.44^{+0.13}_{-0.23}$ & TAP transit model\\
 & Transit duration, $\tau_{\rm{c}}$ & min & $-$ & $771^{+66}_{-40}$ & TAP transit model\\
 & Total transit duration, $\tau_{\rm{14, c}}$ & min & $-$ & $838^{+71}_{-41}$ & derived\\
 & Transit mid-point, $T_{\rm{mid, c}}$ & $\rm BJD_{TDB}$ & $-$ & $2460062.593^{+0.014}_{-0.024}$ & TAP transit model\\
 & Mass, $M_{\rm{c}}$ & $M_{\rm{Jup}}$ & $\mathcal{U}(1,40)$ & $14.2^{+1.0}_{-1.2}$ & TTVFast dynamical model\\
 & Radius, $R_{\rm{c}}$ & $R_{\rm{Jup}}$ & $-$ & $0.993^{+0.017}_{-0.014}$ & derived\\
 & Density, $\rho_{\rm{c}}$ & $\rm{g~cm}^{-3}$ & $-$ & $19.2^{+1.0}_{-1.2}$ & derived\\
 & Semi-major axis, $a_{\rm{c}}$ & au & $-$ & $4.28^{+0.36}_{-0.21}$ & derived\\
 & RV amplitude, $K_{\rm{c}}$ & $\rm{m~s}^{-1}$ & $-$ & $221^{+16}_{-18}$ & derived\\
\multicolumn{6}{l}{Common} \\
 & Mutual orbital inclination, $i_{\rm bc}$ & $^{\circ}$ & $-$ & $2.9^{+4.8}_{-4.4}$ & derived\\
 & Stellar mass, $M_{\star}$ & $M_{\odot}$ & $-$ & $1.316\pm0.027$ & \citet{2021AJ....161..235H}\\
 & Stellar radius, $R_{\star}$ & $R_{\odot}$ & $\mathcal{U}(1.2,1.5)$ & $1.339^{+0.017}_{-0.011}$ & TTVFast dynamical model\\
 & Linear LD coefficient, $u_{\rm 1, TESS}$ & $-$ & $\mathcal{N}(0.23,0.10)$ & $0.211\pm0.023$ & TAP transit model\\
 & Quadratic LD coefficient, $u_{\rm 2, TESS}$ & $-$ & $\mathcal{N}(0.32,0.20)$ & $0.267\pm0.036$ & TAP transit model\\
 & HARPS barycentric RV, $\gamma_{\rm HARPS}$ & $\rm{m~s}^{-1}$ & $\mathcal{U}(16770,17070)$ & $16905^{+11}_{-17}$ & TTVFast dynamical model\\
 & HARPS RV jitter, $\sigma_{\rm jitter, HARPS}$ & $\rm{m~s}^{-1}$ & $\mathcal{U}(0,50)$ & $10.5^{+3.1}_{-1.6}$ & TTVFast dynamical model\\
 & Goodness of fit, $\chi^2$ & $-$ & $-$ & $54.4$ & TTVFast dynamical model\\
 & Number of degrees of freedom, $N_{\rm dof}$ & $-$ & $-$ & $71$ & TTVFast dynamical model\\
\enddata
\tablecomments{Priors are given for free parameters of the dynamical fit and for the LD coefficients in transit light curves modeling. The orbital parameters are given in the astrocentric reference system for epoch $\rm BJD_{TDB}$ 2,458,158.087205.}
\end{deluxetable*}

\begin{deluxetable*}{clccccccccc}
\tablewidth{0pt}
\tablecaption{Parameters for individual transits of TOI-201~b \label{tab:ttimes}}
\tablehead{
\colhead{Sector} & \colhead{Date} & \colhead{$E$} & \colhead{$T_{\rm mid}$ ($\rm BJD_{TDB}$)} & \colhead{$R_{\rm b}/R_{\star}$} & \colhead{$b_{\rm b}/R_{\star}$} & \colhead{$\tau_{\rm b}$ (min)} & \colhead{$\chi^2$} & \colhead{$N_{\rm dof}$} & \colhead{$\hat{R}_{\rm z}$} & \colhead{$N_{\rm ess}$}
}
\startdata
 2 & 2018.7029 &  0 & $2458376.05223^{+0.00062}_{-0.00062}$ & $0.0800^{+0.0010}_{-0.0010}$ & $0.729^{+0.032}_{-0.039}$ & $247.4^{+2.1}_{-2.2}$ & $1081.7$ & $829$ & $1.018$ & $6472$ \\
 4 & 2018.8480 &  1 & $2458429.02932^{+0.00066}_{-0.00067}$ & $0.0780^{+0.0011}_{-0.0011}$ & $0.683^{+0.040}_{-0.054}$ & $250.8^{+2.2}_{-2.1}$ & $1348.3$ & $853$ & $1.016$ & $6593$ \\
 6 & 2018.9932 &  2 & $2458482.00932^{+0.00066}_{-0.00065}$ & $0.0794^{+0.0011}_{-0.0011}$ & $0.733^{+0.031}_{-0.038}$ & $246.3^{+2.3}_{-2.2}$ & $1362.9$ & $855$ & $1.020$ & $6804$ \\
10 & 2019.2835 &  4 & $2458587.96457^{+0.00073}_{-0.00071}$ & $0.0797^{+0.0012}_{-0.0012}$ & $0.728^{+0.035}_{-0.045}$ & $247.1^{+2.4}_{-2.4}$ & $1518.3$ & $854$ & $1.013$ & $6758$ \\
12 & 2019.4286 &  5 & $2458640.94287^{+0.00068}_{-0.00069}$ & $0.0792^{+0.0012}_{-0.0012}$ & $0.691^{+0.039}_{-0.049}$ & $247.6^{+2.3}_{-2.3}$ & $1182.9$ & $803$ & $1.013$ & $6904$ \\
28 & 2020.5882 & 13 & $2459064.76853^{+0.00062}_{-0.00064}$ & $0.0788^{+0.0010}_{-0.0010}$ & $0.745^{+0.028}_{-0.034}$ & $251.1^{+2.1}_{-2.1}$ & $1075.5$ & $854$ & $1.026$ & $6374$ \\
30 & 2020.7329 & 14 & $2459117.74784^{+0.00072}_{-0.00074}$ & $0.0793^{+0.0011}_{-0.0011}$ & $0.748^{+0.026}_{-0.031}$ & $248.3^{+2.5}_{-2.4}$ & $1358.5$ & $854$ & $1.010$ & $7819$ \\
33 & 2021.0225 & 16 & $2459223.70358^{+0.00059}_{-0.00059}$ & $0.0792^{+0.0010}_{-0.0010}$ & $0.719^{+0.033}_{-0.039}$ & $247.4^{+2.0}_{-2.0}$ & $1117.1$ & $854$ & $1.017$ & $6271$ \\
35 & 2021.1676 & 17 & $2459276.68161^{+0.00058}_{-0.00059}$ & $0.0794^{+0.0010}_{-0.0010}$ & $0.737^{+0.028}_{-0.036}$ & $247.2^{+2.1}_{-2.1}$ & $1075.2$ & $855$ & $1.020$ & $6411$ \\
37 & 2021.3128 & 18 & $2459329.65924^{+0.00061}_{-0.00062}$ & $0.0777^{+0.0011}_{-0.0012}$ & $0.647^{+0.050}_{-0.074}$ & $247.6^{+2.0}_{-2.0}$ & $1244.8$ & $854$ & $1.015$ & $6194$ \\
39 & 2021.4579 & 19 & $2459382.63808^{+0.00066}_{-0.00066}$ & $0.0780^{+0.0013}_{-0.0014}$ & $0.671^{+0.054}_{-0.083}$ & $246.4^{+2.2}_{-2.3}$ & $1293.1$ & $840$ & $1.016$ & $5998$ \\
61 & 2023.0545 & 30 & $2459965.39407^{+0.00074}_{-0.00069}$ & $0.0794^{+0.0011}_{-0.0012}$ & $0.740^{+0.027}_{-0.035}$ & $247.1^{+2.4}_{-2.4}$ & $1041.4$ & $855$ & $1.013$ & $7131$ \\
65 & 2023.3448 & 32 & $2460071.36864^{+0.00057}_{-0.00058}$ & $0.0779^{+0.0011}_{-0.0011}$ & $0.691^{+0.043}_{-0.052}$ & $247.5^{+1.9}_{-1.9}$ & $890.8$ & $855$ & $1.035$ & $5897$ \\
68 & 2023.6352 & 34 & $2460177.33151^{+0.00060}_{-0.00062}$ & $0.0796^{+0.0010}_{-0.0010}$ & $0.734^{+0.028}_{-0.033}$ & $247.0^{+2.1}_{-2.1}$ & $1115.8$ & $854$ & $1.017$ & $6942$ \\
88 & 2025.0839 & 44 & $2460707.11204^{+0.00065}_{-0.00063}$ & $0.0780^{+0.0011}_{-0.0012}$ & $0.692^{+0.035}_{-0.048}$ & $247.0^{+2.1}_{-2.1}$ & $1069.8$ & $855$ & $1.012$ & $6811$ \\
\enddata
\tablecomments{Date is a calendar year in decimal format. $E$ is a transit number counted from the reference epoch given by Eq.~\ref{Eq:ephem}. $N_{\rm dof}$ in the number of degrees of freedom. $\hat{R}_{\rm z}$ is the multivariate
Gelman-Rubin statistic. $N_{\rm ess}$ is the number of effective samples.}
\end{deluxetable*}

\begin{acknowledgments}
We thank the anonymous referee for their constructive and insightful comments, which helped improve the quality and clarity of this Letter. GM acknowledges the financial support from the National Science Centre, Poland, through grant no. 2023/49/B/ST9/00285. This paper includes data collected with the TESS mission, obtained from the MAST data archive at the Space Telescope Science Institute (STScI). These data can be accessed at \dataset[doi: 10.17909/t9-yk4w-zc73]{\doi{10.17909/t9-yk4w-zc73}} \citep{https://doi.org/10.17909/t9-yk4w-zc73}. Funding for the TESS mission is provided by the NASA Explorer Program. STScI is operated by the Association of Universities for Research in Astronomy, Inc., under NASA contract NAS 5-26555. This research made use of \texttt{Lightkurve}, a Python package for Kepler and TESS data analysis \citep{2018ascl.soft12013L}. This research has made use of the SIMBAD database and the VizieR catalogue access tool, operated at CDS, Strasbourg, France, and NASA's Astrophysics Data System Bibliographic Services.
\end{acknowledgments}

\begin{contribution}
%%This section gives authors the space to recognize author contributions. The text inside this environment is NOT counted towards the total word quanta. At a minimum, manuscripts are expected to include this text:

W\L{} reduced the photometric observations from Sectors 1--68 and conducted the preliminary TTV analysis. GM carried out the remaining research work, including further data analysis and interpretation.

%% But authors are expected to provide more specific details, e.g. 
%%
%%SC was responsible for writing and submitting the manuscript.
%%WWM came up with the initial research concept and edited the manuscript.
%%OTS obtained the funding and edited the manuscript.
%%EBF provided the formal analysis and validation. He also edited the manuscript.
%%GEH Supervised the undergraduates, wrote the software and administers the project github and Zenodo repositories.
%%
%% Authors can use the Contributor Role Taxonomy (CRediT) at
%% for ideas on how write a good statement tailored to their needs.

\end{contribution}

\facility{TESS}

\software{Lightkurve \citep{2018ascl.soft12013L},
          Transit Analysis Package \citep{2012AdAst2012E..30G}, 
          TTVFast \citep{2014ApJ...787..132D},
          dynesty \citep{2020MNRAS.493.3132S},
          SWIFT \citep{2013ascl.soft03001L}
          }


%\appendix

%% For this sample we use BibTeX plus aasjournalv7.bst to generate the

\bibliography{toi201}{}

\begin{thebibliography}{}
\expandafter\ifx\csname natexlab\endcsname\relax\def\natexlab#1{#1}\fi
\providecommand{\url}[1]{\href{#1}{#1}}
\providecommand{\dodoi}[1]{doi:~\href{http://doi.org/#1}{\nolinkurl{#1}}}
\providecommand{\doeprint}[1]{\href{http://ascl.net/#1}{\nolinkurl{http://ascl.net/#1}}}
\providecommand{\doarXiv}[1]{\href{https://arxiv.org/abs/#1}{\nolinkurl{https://arxiv.org/abs/#1}}}

\bibitem[{E. {Agol} {et~al.}(2021){Agol}, {Dorn}, {Grimm}, {Turbet}, {Ducrot}, {Delrez}, {Gillon}, {Demory}, {Burdanov}, {Barkaoui}, {Benkhaldoun}, {Bolmont}, {Burgasser}, {Carey}, {de Wit}, {Fabrycky}, {Foreman-Mackey}, {Haldemann}, {Hernandez}, {Ingalls}, {Jehin}, {Langford}, {Leconte}, {Lederer}, {Luger}, {Malhotra}, {Meadows}, {Morris}, {Pozuelos}, {Queloz}, {Raymond}, {Selsis}, {Sestovic}, {Triaud}, \& {Van Grootel}}]{2021PSJ.....2....1A}
{Agol}, E., {Dorn}, C., {Grimm}, S.~L., {et~al.} 2021, \bibinfo{title}{{Refining the Transit-timing and Photometric Analysis of TRAPPIST-1: Masses, Radii, Densities, Dynamics, and Ephemerides},} \psj, 2, 1, \dodoi{10.3847/PSJ/abd022}

\bibitem[{A. {Claret} \& S. {Bloemen}(2011){Claret} \& {Bloemen}}]{2011AA...529A..75C}
{Claret}, A., \& {Bloemen}, S. 2011, \bibinfo{title}{{Gravity and limb-darkening coefficients for the Kepler, CoRoT, Spitzer, uvby, UBVRIJHK, and Sloan photometric systems},} \aap, 529, A75, \dodoi{10.1051/0004-6361/201116451}

\bibitem[{A.~C.~M. {Correia} {et~al.}(2005){Correia}, {Udry}, {Mayor}, {Laskar}, {Naef}, {Pepe}, {Queloz}, \& {Santos}}]{2005A&A...440..751C}
{Correia}, A.~C.~M., {Udry}, S., {Mayor}, M., {et~al.} 2005, \bibinfo{title}{{The CORALIE survey for southern extra-solar planets. XIII. A pair of planets around HD{\,}202206 or a circumbinary planet?},} \aap, 440, 751, \dodoi{10.1051/0004-6361:20042376}

\bibitem[{N.~B. {Cowan} \& E. {Agol}(2011){Cowan} \& {Agol}}]{2011ApJ...726...82C}
{Cowan}, N.~B., \& {Agol}, E. 2011, \bibinfo{title}{{A Model for Thermal Phase Variations of Circular and Eccentric Exoplanets},} \apj, 726, 82, \dodoi{10.1088/0004-637X/726/2/82}

\bibitem[{K.~M. {Deck} {et~al.}(2014){Deck}, {Agol}, {Holman}, \& {Nesvorn{\'y}}}]{2014ApJ...787..132D}
{Deck}, K.~M., {Agol}, E., {Holman}, M.~J., \& {Nesvorn{\'y}}, D. 2014, \bibinfo{title}{{TTVFast: An Efficient and Accurate Code for Transit Timing Inversion Problems},} \apj, 787, 132, \dodoi{10.1088/0004-637X/787/2/132}

\bibitem[{J. {Dong} {et~al.}(2021){Dong}, {Huang}, {Dawson}, {Foreman-Mackey}, {Collins}, {Quinn}, {Lissauer}, {Beatty}, {Quarles}, {Sha}, {Shporer}, {Guo}, {Kane}, {Abe}, {Barkaoui}, {Benkhaldoun}, {Brahm}, {Bouchy}, {Carmichael}, {Collins}, {Conti}, {Crouzet}, {Dransfield}, {Evans}, {Gan}, {Ghachoui}, {Gillon}, {Grieves}, {Guillot}, {Hellier}, {Jehin}, {Jensen}, {Jord{\'a}n}, {Kamler}, {Kielkopf}, {M{\'e}karnia}, {Nielsen}, {Pozuelos}, {Radford}, {Schmider}, {Schwarz}, {Stockdale}, {Tan}, {Timmermans}, {Triaud}, {Wang}, {Ricker}, {Vanderspek}, {Latham}, {Seager}, {Winn}, {Jenkins}, {Mireles}, {Yahalomi}, {Morgan}, {Vezie}, {Quintana}, {Rose}, {Smith}, \& {Shiao}}]{2021ApJS..255....6D}
{Dong}, J., {Huang}, C.~X., {Dawson}, R.~I., {et~al.} 2021, \bibinfo{title}{{Warm Jupiters in TESS Full-frame Images: A Catalog and Observed Eccentricity Distribution for Year 1},} \apjs, 255, 6, \dodoi{10.3847/1538-4365/abf73c}

\bibitem[{J.~Z. {Gazak} {et~al.}(2012){Gazak}, {Johnson}, {Tonry}, {Dragomir}, {Eastman}, {Mann}, \& {Agol}}]{2012AdAst2012E..30G}
{Gazak}, J.~Z., {Johnson}, J.~A., {Tonry}, J., {et~al.} 2012, \bibinfo{title}{{Transit Analysis Package: An IDL Graphical User Interface for Exoplanet Transit Photometry},} Advances in Astronomy, 2012, 697967, \dodoi{10.1155/2012/697967}

\bibitem[{H.~A.~C. {Giles} {et~al.}(2018){Giles}, {Osborn}, {Blanco-Cuaresma}, {Lovis}, {Bayliss}, {Eggenberger}, {Collier Cameron}, {Kristiansen}, {Turner}, {Bouchy}, \& {Udry}}]{2018A&A...615L..13G}
{Giles}, H.~A.~C., {Osborn}, H.~P., {Blanco-Cuaresma}, S., {et~al.} 2018, \bibinfo{title}{{Transiting planet candidate from K2 with the longest period},} \aap, 615, L13, \dodoi{10.1051/0004-6361/201833569}

\bibitem[{N.~M. {Guerrero} {et~al.}(2021){Guerrero}, {Seager}, {Huang}, {Vanderburg}, {Garcia Soto}, {Mireles}, {Hesse}, {Fong}, {Glidden}, {Shporer}, {Latham}, {Collins}, {Quinn}, {Burt}, {Dragomir}, {Crossfield}, {Vanderspek}, {Fausnaugh}, {Burke}, {Ricker}, {Daylan}, {Essack}, {G{\"u}nther}, {Osborn}, {Pepper}, {Rowden}, {Sha}, {Villanueva}, {Yahalomi}, {Yu}, {Ballard}, {Batalha}, {Berardo}, {Chontos}, {Dittmann}, {Esquerdo}, {Mikal-Evans}, {Jayaraman}, {Krishnamurthy}, {Louie}, {Mehrle}, {Niraula}, {Rackham}, {Rodriguez}, {Rowden}, {Sousa-Silva}, {Watanabe}, {Wong}, {Zhan}, {Zivanovic}, {Christiansen}, {Ciardi}, {Swain}, {Lund}, {Mullally}, {Fleming}, {Rodriguez}, {Boyd}, {Quintana}, {Barclay}, {Col{\'o}n}, {Rinehart}, {Schlieder}, {Clampin}, {Jenkins}, {Twicken}, {Caldwell}, {Coughlin}, {Henze}, {Lissauer}, {Morris}, {Rose}, {Smith}, {Tenenbaum}, {Ting}, {Wohler}, {Bakos}, {Bean}, {Berta-Thompson}, {Bieryla}, {Bouma}, {Buchhave}, {Butler}, {Charbonneau}, {Doty}, {Ge}, {Holman}, {Howard}, {Kaltenegger},
  {Kane}, {Kjeldsen}, {Kreidberg}, {Lin}, {Minsky}, {Narita}, {Paegert}, {P{\'a}l}, {Palle}, {Sasselov}, {Spencer}, {Sozzetti}, {Stassun}, {Torres}, {Udry}, \& {Winn}}]{2021ApJS..254...39G}
{Guerrero}, N.~M., {Seager}, S., {Huang}, C.~X., {et~al.} 2021, \bibinfo{title}{{The TESS Objects of Interest Catalog from the TESS Prime Mission},} \apjs, 254, 39, \dodoi{10.3847/1538-4365/abefe1}

\bibitem[{A.~P. {Hatzes} \& H. {Rauer}(2015){Hatzes} \& {Rauer}}]{2015ApJ...810L..25H}
{Hatzes}, A.~P., \& {Rauer}, H. 2015, \bibinfo{title}{{A Definition for Giant Planets Based on the Mass-Density Relationship},} \apjl, 810, L25, \dodoi{10.1088/2041-8205/810/2/L25}

\bibitem[{M.~J. {Hobson} {et~al.}(2021){Hobson}, {Brahm}, {Jord{\'a}n}, {Espinoza}, {Kossakowski}, {Henning}, {Rojas}, {Schlecker}, {Sarkis}, {Trifonov}, {Thorngren}, {Binnenfeld}, {Shahaf}, {Zucker}, {Ricker}, {Latham}, {Seager}, {Winn}, {Jenkins}, {Addison}, {Bouchy}, {Bowler}, {Briegal}, {Bryant}, {Collins}, {Daylan}, {Grieves}, {Horner}, {Huang}, {Kane}, {Kielkopf}, {McLean}, {Mengel}, {Nielsen}, {Okumura}, {Jones}, {Plavchan}, {Shporer}, {Smith}, {Tilbrook}, {Tinney}, {Twicken}, {Udry}, {Unger}, {West}, {Wittenmyer}, {Wohler}, {Torres}, \& {Wright}}]{2021AJ....161..235H}
{Hobson}, M.~J., {Brahm}, R., {Jord{\'a}n}, A., {et~al.} 2021, \bibinfo{title}{{A Transiting Warm Giant Planet around the Young Active Star TOI-201},} \aj, 161, 235, \dodoi{10.3847/1538-3881/abeaa1}

\bibitem[{N. {Houk} \& A.~P. {Cowley}(1975){Houk} \& {Cowley}}]{1975mcts.book.....H}
{Houk}, N., \& {Cowley}, A.~P. 1975, {University of Michigan Catalogue of two-dimensional spectral types for the HD stars. Volume I. Declinations -90\_ to -53\_{\textflorin}0.}

\bibitem[{D. {Kipping}(2021){Kipping}}]{2021MNRAS.500.1851K}
{Kipping}, D. 2021, \bibinfo{title}{{The exomoon corridor: Half of all exomoons exhibit TTV frequencies within a narrow window due to aliasing},} \mnras, 500, 1851, \dodoi{10.1093/mnras/staa3398}

\bibitem[{R.~K. {Kopparapu} {et~al.}(2013){Kopparapu}, {Ramirez}, {Kasting}, {Eymet}, {Robinson}, {Mahadevan}, {Terrien}, {Domagal-Goldman}, {Meadows}, \& {Deshpande}}]{2013ApJ...765..131K}
{Kopparapu}, R.~K., {Ramirez}, R., {Kasting}, J.~F., {et~al.} 2013, \bibinfo{title}{{Habitable Zones around Main-sequence Stars: New Estimates},} \apj, 765, 131, \dodoi{10.1088/0004-637X/765/2/131}

\bibitem[{J. {Laskar}(1990){Laskar}}]{1990Icar...88..266L}
{Laskar}, J. 1990, \bibinfo{title}{{The chaotic motion of the solar system: A numerical estimate of the size of the chaotic zones},} \icarus, 88, 266, \dodoi{10.1016/0019-1035(90)90084-M}

\bibitem[{J. {Laskar}(1993){Laskar}}]{1993PhyD...67..257L}
{Laskar}, J. 1993, \bibinfo{title}{{Frequency analysis for multi-dimensional systems. Global dynamics and diffusion},} Physica D Nonlinear Phenomena, 67, 257, \dodoi{10.1016/0167-2789(93)90210-R}

\bibitem[{H.~F. {Levison} \& M.~J. {Duncan}(1994){Levison} \& {Duncan}}]{1994Icar..108...18L}
{Levison}, H.~F., \& {Duncan}, M.~J. 1994, \bibinfo{title}{{The Long-Term Dynamical Behavior of Short-Period Comets},} \icarus, 108, 18, \dodoi{10.1006/icar.1994.1039}

\bibitem[{H.~F. {Levison} \& M.~J. {Duncan}(2013){Levison} \& {Duncan}}]{2013ascl.soft03001L}
{Levison}, H.~F., \& {Duncan}, M.~J. 2013, \bibinfo{title}{{SWIFT: A solar system integration software package},}, Astrophysics Source Code Library, record ascl:1303.001

\bibitem[{ {Lightkurve Collaboration} {et~al.}(2018){Lightkurve Collaboration}, {Cardoso}, {Hedges}, {Gully-Santiago}, {Saunders}, {Cody}, {Barclay}, {Hall}, {Sagear}, {Turtelboom}, {Zhang}, {Tzanidakis}, {Mighell}, {Coughlin}, {Bell}, {Berta-Thompson}, {Williams}, {Dotson}, \& {Barentsen}}]{2018ascl.soft12013L}
{Lightkurve Collaboration}, {Cardoso}, J. V. d.~M., {Hedges}, C., {et~al.} 2018, \bibinfo{title}{{Lightkurve: Kepler and TESS time series analysis in Python},} \doeprint{1812.013}

\bibitem[{W. {\L{}oboda}(2024){\L{}oboda}}]{2024Loboda}
{\L{}oboda}, W. 2024, Master's thesis, Nicolaus Copernicus University, Toru\'n, Poland.
\newblock \url{https://apd.umk.pl/diplomas/148300/}

\bibitem[{M.~S. {Marley} {et~al.}(1999){Marley}, {Gelino}, {Stephens}, {Lunine}, \& {Freedman}}]{1999ApJ...513..879M}
{Marley}, M.~S., {Gelino}, C., {Stephens}, D., {Lunine}, J.~I., \& {Freedman}, R. 1999, \bibinfo{title}{{Reflected Spectra and Albedos of Extrasolar Giant Planets. I. Clear and Cloudy Atmospheres},} \apj, 513, 879, \dodoi{10.1086/306881}

\bibitem[{ {MAST Team}(2021){MAST Team}}]{https://doi.org/10.17909/t9-yk4w-zc73}
{MAST Team}. 2021, \bibinfo{title}{TESS Target Pixel Files - All Sectors,} STScI/MAST, \dodoi{10.17909/T9-YK4W-ZC73}

\bibitem[{M. {Montalto} {et~al.}(2020){Montalto}, {Borsato}, {Granata}, {Lacedelli}, {Malavolta}, {Manthopoulou}, {Nardiello}, {Nascimbeni}, \& {Piotto}}]{2020MNRAS.498.1726M}
{Montalto}, M., {Borsato}, L., {Granata}, V., {et~al.} 2020, \bibinfo{title}{{A search for transiting planets around FGKM dwarfs and subgiants in the TESS full frame images of the Southern ecliptic hemisphere},} \mnras, 498, 1726, \dodoi{10.1093/mnras/staa2438}

\bibitem[{S. {M{\"u}ller} {et~al.}(2024){M{\"u}ller}, {Baron}, {Helled}, {Bouchy}, \& {Parc}}]{2024A&A...686A.296M}
{M{\"u}ller}, S., {Baron}, J., {Helled}, R., {Bouchy}, F., \& {Parc}, L. 2024, \bibinfo{title}{{The mass-radius relation of exoplanets revisited},} \aap, 686, A296, \dodoi{10.1051/0004-6361/202348690}

\bibitem[{J.~A. {Patel} \& N. {Espinoza}(2022){Patel} \& {Espinoza}}]{2022AJ....163..228P}
{Patel}, J.~A., \& {Espinoza}, N. 2022, \bibinfo{title}{{Empirical Limb-darkening Coefficients and Transit Parameters of Known Exoplanets from TESS},} \aj, 163, 228, \dodoi{10.3847/1538-3881/ac5f55}

\bibitem[{G.~R. {Ricker} {et~al.}(2015){Ricker}, {Winn}, {Vanderspek}, {Latham}, {Bakos}, {Bean}, {Berta-Thompson}, {Brown}, {Buchhave}, {Butler}, {Butler}, {Chaplin}, {Charbonneau}, {Christensen-Dalsgaard}, {Clampin}, {Deming}, {Doty}, {De Lee}, {Dressing}, {Dunham}, {Endl}, {Fressin}, {Ge}, {Henning}, {Holman}, {Howard}, {Ida}, {Jenkins}, {Jernigan}, {Johnson}, {Kaltenegger}, {Kawai}, {Kjeldsen}, {Laughlin}, {Levine}, {Lin}, {Lissauer}, {MacQueen}, {Marcy}, {McCullough}, {Morton}, {Narita}, {Paegert}, {Palle}, {Pepe}, {Pepper}, {Quirrenbach}, {Rinehart}, {Sasselov}, {Sato}, {Seager}, {Sozzetti}, {Stassun}, {Sullivan}, {Szentgyorgyi}, {Torres}, {Udry}, \& {Villasenor}}]{2015JATIS...1a4003R}
{Ricker}, G.~R., {Winn}, J.~N., {Vanderspek}, R., {et~al.} 2015, \bibinfo{title}{{Transiting Exoplanet Survey Satellite (TESS)},} Journal of Astronomical Telescopes, Instruments, and Systems, 1, 014003, \dodoi{10.1117/1.JATIS.1.1.014003}

\bibitem[{A. {Savitzky} \& M.~J.~E. {Golay}(1964){Savitzky} \& {Golay}}]{1964AnaCh..36.1627S}
{Savitzky}, A., \& {Golay}, M.~J.~E. 1964, \bibinfo{title}{{Smoothing and differentiation of data by simplified least squares procedures},} Analytical Chemistry, 36, 1627, \dodoi{10.1021/ac60214a047}

\bibitem[{K.~C. {Schlaufman}(2018){Schlaufman}}]{2018ApJ...853...37S}
{Schlaufman}, K.~C. 2018, \bibinfo{title}{{Evidence of an Upper Bound on the Masses of Planets and Its Implications for Giant Planet Formation},} \apj, 853, 37, \dodoi{10.3847/1538-4357/aa961c}

\bibitem[{J.~S. {Speagle}(2020){Speagle}}]{2020MNRAS.493.3132S}
{Speagle}, J.~S. 2020, \bibinfo{title}{{DYNESTY: a dynamic nested sampling package for estimating Bayesian posteriors and evidences},} \mnras, 493, 3132, \dodoi{10.1093/mnras/staa278}

\end{thebibliography}
\bibliographystyle{aasjournalv7}



\end{document}